# Symmetric reflection line resonator for semiconductor circuit quantum electrodynamics

Miao-Lei Zhang[1], Guang-Wei Deng[1], Shu-Xiao Li[1], Hai-Ou Li[1], Gang Cao[1], Tao Tu[1], Ming Xiao[1], Guang-Can Guo[1], Hong-Wen Jiang[2], Irfan Siddiqi[3] and Guo-Ping Guo[1*]

[1] Key Laboratory of Quantum Information, University of Science and Technology of China, Hefei 230026, People's Republic of China

[2] Department of Physics and Astronomy, University of California at Los Angeles, California 90095, USA.

[3] Quantum Nanoelectronics Laboratory, Department of Physics, University of California, Berkeley, CA 94720, USA.

**We have designed and fabricated a half-wavelength reflection line resonator (RLR) that consists of a pair of two coupled microstrip lines on a GaAs/AlGaAs heterostructure. By changing the top gate voltage on a square of two dimensional electron gas under the resonator, a large range of the quality factors can be obtained. Energy loss in the two-dimensional electron gas can be minimized, thus realizing a versatile resonator suitable for integration with semiconductor quantum circuits.**

Superconducting transmission line resonators (TLR) based on coplanar waveguide (CPW) have played an important role in the circuit quantum electrodynamics architecture[1-4]. They can be used for single qubit control[5,6], for dispersive qubit readout[7,8], and for coupling multiple qubits as a quantum bus[9-11]. Very recently, a promise qubit candidate system gate confined quantum dots[12-14], have been coupled to a TLR[15,16]. However, in typical semiconductor devices, the confinement gates of the quantum dots couple to a large area of the CPW ground plane, potentially resulting in extra energy leakage. A resonator design with no ground plane can avoid this problem. In this letter, we present a half-wavelength on-chip cavity on doped GaAs that consists of a pair of coupled microstrip lines. We call this structure a reflection line resonator (RLR) as we measure the reflected signal off the cavity instead of a transmission response in a TLR. Our design is similar to that developed in UC Berkely[17]. There is no ground plane and the two ends of our resonator are both free to couple qubits via a symmetric, differential excitation, which can potentially have a larger coupling strength compared with a TLR which offers only one center pin for coupling and immunity to common-mode noise. The resonator quality factor can be varied by way of a gate bias and low loss operation was achieved.

We studied the properties of coupling between our resonator and the GaAs/AlGaAs two-dimensional electron gas (2DEG). By controlling the density of a 2DEG coupled to the resonator, a large range of quality factor tuning is observed. This allows us to assess the loss contribution of the 2DEG layer and assess what range of quality factors can be practically achieved.

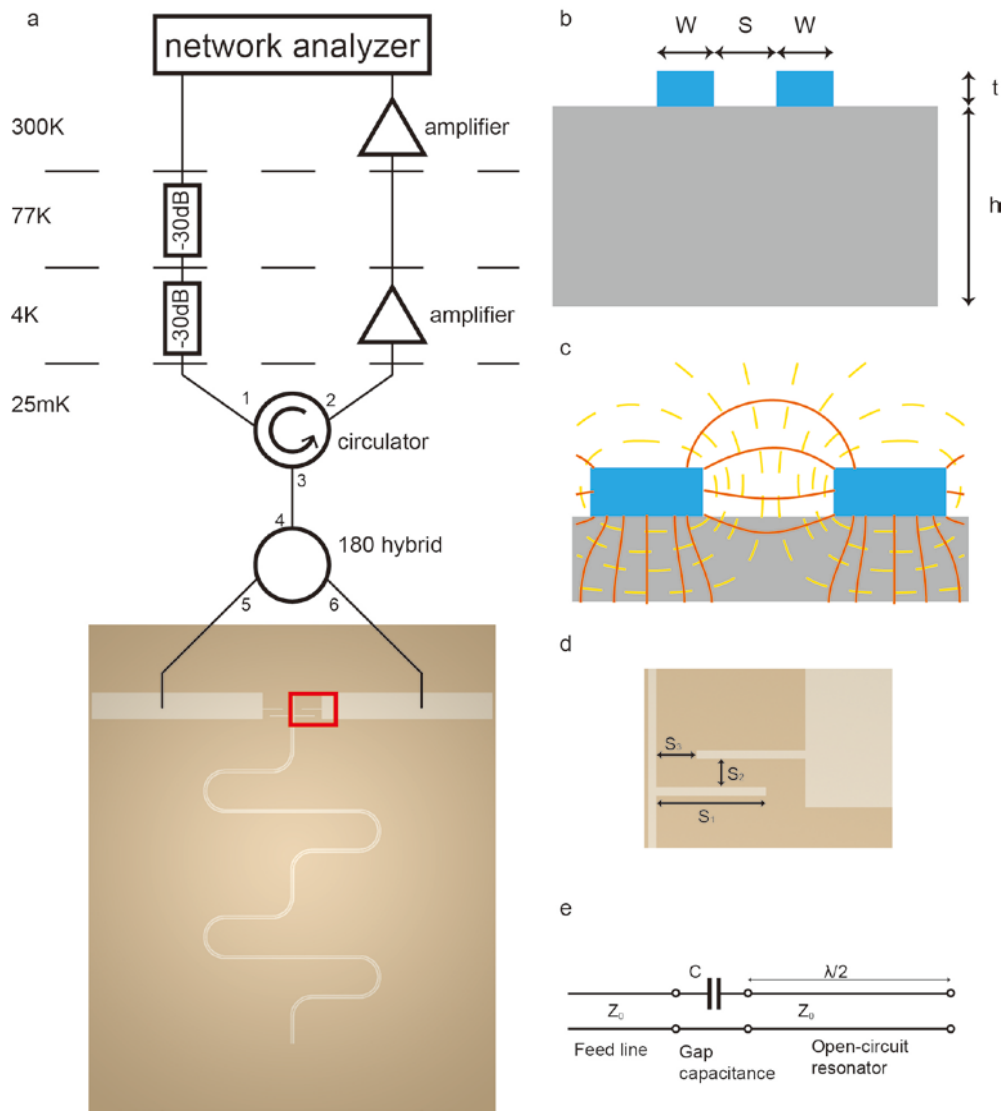

FIG 1 (Color online) (a) Optical microscope image of the reflection line resonator and schematic of the measurement setup. (b) Cross section of the coupled microstrip lines: two parallel conductor metallization on top of the GaAs substrate. (c) Schematic drawing of the electric and magnetic field lines for the fundamental mode of coupled microstrip lines. (d) Magnified view of a finger capacitors, location on the marked rectangle in (a). (e) Equivalent circuit model of coupled microstrip reflection line resonator with gap.

The sample is shown in fig 1(a) along with its measurement circuit schematic. The resonator is formed by a pair of coupled microstrip lines of width w=9um separated by a gap of width s=16um. The length of the resonator is designed with l=9mm to obtain a resonant frequency of 6GHz. These center conductors are coupled via finger capacitors to the input/output transmission lines. The finger capacitors formed by one pair of fingers of length s1=100um, and separation s2=30um, s3=50um were designed and fabricated, see fig 1(d).

The resonator was fabricated on a GaAs/AlGaAs heterostructure crystal with a 2DEG about 100 nm deep below the surface. To avoid energy leakage from the substrate and get high quality factor, the 2DEG was removed with a 200nm-deep wet etch. The total thickness of the substrate is 650um. The resonator was patterned using optical lithography with a 2um thick layer of photoresist. The wafer was subsequently metallized with a 300 nm thick layer of Al, electron beam evaporated at 10A/s rate and lifted-off in acetone. A cross-section sketch of the coupled microstrip lines is shown in fig 1(b). We simulated the characteristic electromagnetic field pattern in the coupled microstrip line structure. If we apply equal amplitude but opposite phase microwave power on the two arms of the coupled conductor line simultaneously, a stable electromagnetic field could be established between them, as shown in the fig 1(c). It's the fundamental mode of the coupled microstrip line. In this case, if the length of the couple lines coincides with half of the wavelength of the input microwave excitation, resonance behavior will be observed.

Using a network analyzer, we measured the reflection spectrum of the GaAs sample, as described in fig 1(a). Two key devices in the measurement circuit are a circulator and a 180 degree hybrid. An ideal circulator is a non-reciprocal three port device in which the microwaves can only be transmitted along one direction: a signal applied to port 1 only comes out of port 2; a signal applied to port 2 only comes out of port 3. Using the circulator, reflected signals and input signals could be effectively isolated and the signal to noise ratio was improved. The 180 degree hybrid is a network with a 180 degree phase shift between the two output ports. If the input was applied to port 4, it would be equally split into two components with a 180 degree phase difference at ports 5 and 6. Conversely, if two signals were applied at ports 5 and 6 respectively, the difference of the inputs would be formed at port 4. Thus, the 180 degree hybrid's function was to change the single-ended input microwave to a differential excitation combining the two reflection signals to one. The final reflection signal was magnified by a 32dB gain HEMT amplifier at 4 K as well as one room temperature amplifier with 35 dB gain. The power applied to the resonator was below -110 dBm to avoid thermal or nonlinear effects. All the measurements were made in a dilution refrigerator with a base temperature of 30mK.

The reflection spectrum of the resonator is shown in fig 2. Near resonance, the magnitude of the reflection signal diminished and a phase reversal is observed. To extract the quality factor, we fit the phase data by using $\lambda/2$ open-circuited microstrip resonator model[18], which gives

$$S_{21} = \frac{1 - g + j*2Q_{int}\frac{\Delta\omega}{\omega_0}}{1 + g + j*2Q_{int}\frac{\Delta\omega}{\omega_0}} \quad (1)$$

With

$$g = Q_{int}/Q_{ext} \quad (2)$$

The cavity transmission and phase shift are determined by equation(1) with $A = |S_{21}|$ and $\phi = \arg(S_{21})$. By fitting the data, we can get internal quality factor $Q_{int} = 2696$ and the external quality factor $Q_{ext} = 4381$. We notice that in our reflection line resonator the $Q_{int}$ and $Q_{ext}$ are naturally obtained together in one step of data fitting[19].

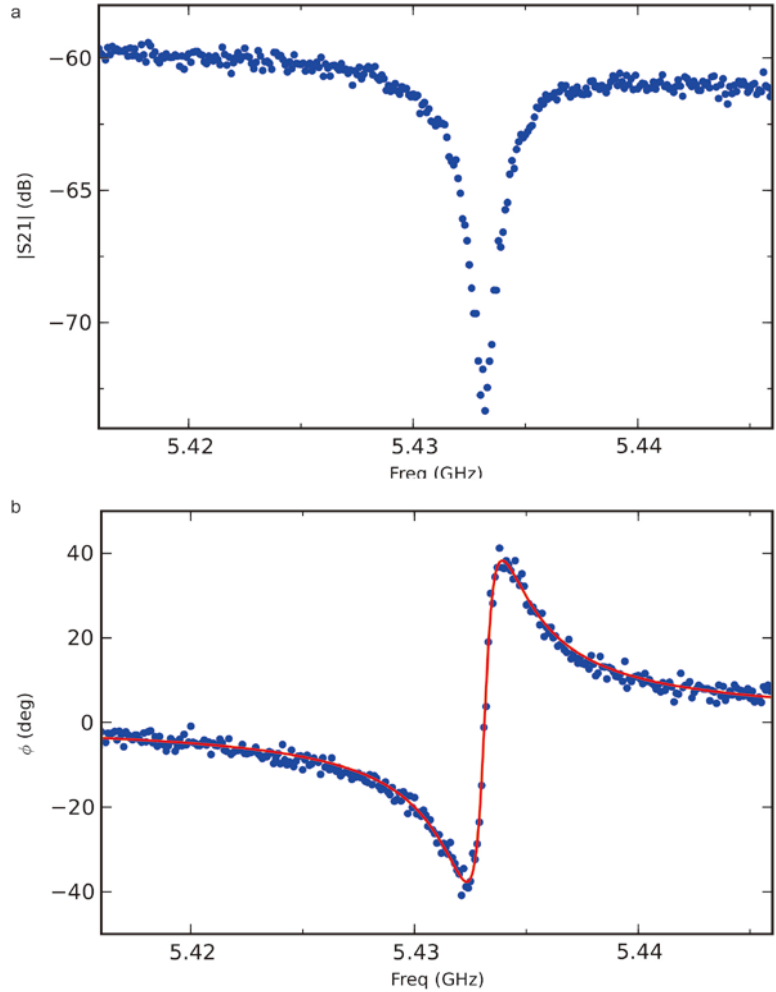

FIG.2. Measured resonator reflection spectra (a) Magnitude response of the cavity as a function of frequency. (b)

Measured phase spectra data points (blue) and the fitting results (red)

$Q_{int}$ and $Q_{ext}$ characterize the internal and external energy loss respectively. The internal energy loss is determined by the resonator chip itself, which mainly includes resistive loss, radiation loss and dielectric loss. The resistive conductor loss of the resonator strips be small since aluminum becomes superconductive at base temperature. Radiation loss is also presumably negligible due to the small separation between the two lines of the resonator. Thus, we suspect dielectric loss to be the main source of energy leakage in our sample, likely due to the finite resistivity of the GaAs substrate which is not as high as sapphire or intrinsic silicon. The external energy loss results from the attached measurement circuit, which depends on the coupling capacitance. The loaded Q is

$$\frac{1}{Q_L} = \frac{1}{Q_{int}} + \frac{1}{Q_{ext}} \qquad (3)$$

In the next step of our experiment, we designed a new architecture to adjust the leakage rate of the resonator. As shown in fig 3(a), a small square of the GaAs/AlGaAs heterostructure is reserved during wet etching. After the lift-off of the aluminum resonator, a 50nm thick of PMMA were over-exposure to form an insulation layer above the resonator. Then another layer of 60 nm thick Al was deposited as the top gate.

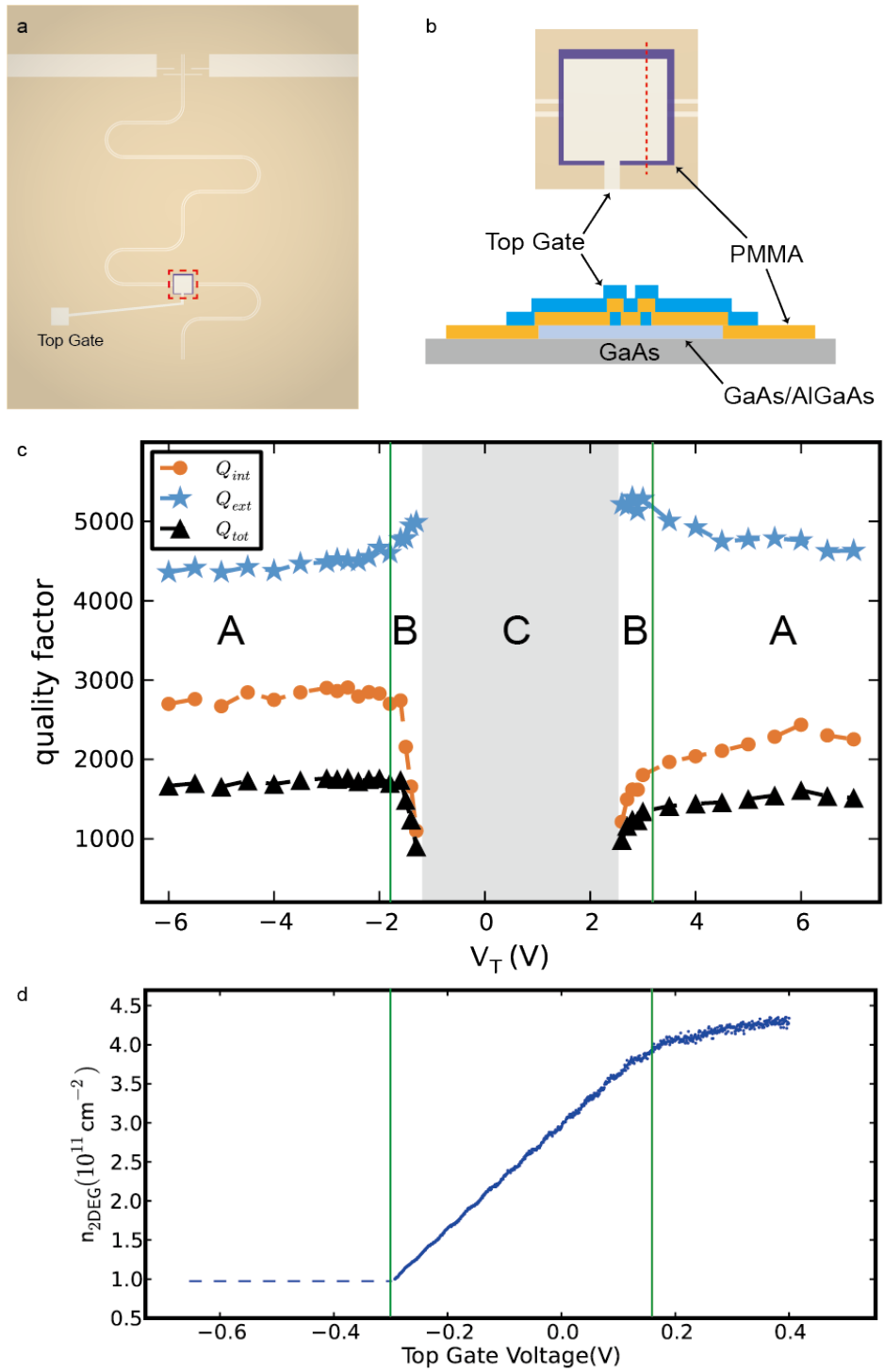

FIG.3. (Color online) (a) Optical microscope image of the reflection line resonator coupled to 2DEG. (b) Cross section of the resonator with 2DEG, PMMA and top gate. (c) Internal quality factor (orange point), external quality factor (blue star) and total quality factor (black triangle) in three voltage intervals: (A) steady regime, (B) tunable regime, (C) vanishing regime. (d) Typical electron density-top gate voltage curve of the sample wafer in Hall bar structure without PMMA.

We studied the effect of the top gate voltage on the microwave properties of the cavity at the base temperature. At large bias voltage $V_T$ (both positive and negative), a clear resonance was observed, and its properties were not sensitive with varying of the voltage. However, when the voltage became small, the resonance frequency was shifted and the quality factor dropped quickly, even vanished at some voltage range. We fitted the reflection phase data under different bias voltages and show the changes of $Q_{int}$ and $Q_{ext}$ in fig 3(c). There are three different intervals: steady regime where the resonance was explicitly visible, tunable regime where the resonance varied and vanishing regime where the resonance disappeared. $Q_{ext}$ did not significantly change as the $V_T$ changing at the steady regime and got a little larger at the tunable regime. But $Q_{int}$ behaves differently. At the steady regime $Q_{int}$ was stationary, while at the tunable regimes $Q_{int}$ decreased quickly from 2700 to below 1000 as we reduced $V_T$. This could be considered that the change of the resonator's properties is caused by the internal energy leakage but not the external circuit, and at the vanishing regime the extreme internal leakage leads to the failure of resonance.

This behavior can be explained by the existence of the 2DEG. Because the distance of the 2DEG below the substrate surface is much shorter than the separation of the two lines, the free electrons in 2DEG would provide a conductive channel for the high frequency oscillating charge in the resonator, which raised the dielectric loss and deteriorated the internal quality factor. The negative voltage on the top gate could deplete the electron gas in the heterostructure. When the top gate is tuned to some negative voltage, 2DEG density in the heterostuctrue is reduced to zero and $Q_{int}$ became large and stable, as in the left A regime of fig 3(c). When the top gate is tuned to some positive voltage, 2DEG density becomes fixed and $Q_{int}$ becomes also large and stable, as in the right A regime of fig 3(c). We speculated that in this case, the conduction band of the GaAs/AlGaAs heterostructure is fully populated and these electrons in 2DEG cannot absorb the cavity photon so the dielectric loss becomes as small as there is no 2DEG. Between these two regimes of fixed and zero density 2DEG, the free electrons can acts photon absorbers to greatly reduce $Q_{int}$ as in regime B of fig3(c). Tuning the photon absorbing rate of 2DEG can change the cavity internal quality factor. Regime C corresponds to the case where the energy loss is too large to get a measurable $Q_{int}$. Fig 3(d) shows the tuning curve of 2DEG density and the top gate voltage of a Hall bar from the same wafer. As there is no PMMA in this Hall bar, the voltage values in fig 3(d) are about 7 times smaller than that in fig 3(c).

In conclusion, we have designed and fabricated a differential reflection line resonator

(RLR). The resonator properties are consistent with a half-wavelength open-circuited resonator model. Compared with transmission line resonator, our RLR has no ground plane and offers larger space to couple other systems at both ends. By coupling with a square of 2DEG, the quality factor can be largely tunable by the changing of absorbing rate of 2DEG. In regimes of high substrate resistivity, high internal quality factor was obtained a necessity for high fidelity readout of quantum circuits.